\documentclass[twocolumn,showpacs,preprintnumbers,amsmath,amssymb]{revtex4}
\usepackage{graphicx}% Include figure files
\usepackage{dcolumn}% Align table columns on decimal point
\usepackage{bm}% bold math

\begin{document}

\preprint{APS/123-QED}
\title{General stability criterion of two-dimensional inviscid parallel flow }
\author{Liang Sun}
\email{sunl@ustc.edu.cn; sunl@ustc.edu} \affiliation{Dept. of
Modern Mechanics, and School of Earth and Space Sciences,
% University of Science and Technology of China, Hefei, 230026, P.R.China.
\\
 University of Science and
Technology of China, Hefei, 230026, China.}

\date{\today}
\begin{abstract}
  A more restrictively general stability criterion of two-dimensional
inviscid parallel flow is obtained analytically. First, a
sufficient criterion for stability is found as either
$-\mu_1<\frac{U''}{U-U_s}<0$ or $0<\frac{U''}{U-U_s}$ in the flow,
where $U_s$ is the velocity at inflection point, $\mu_1$ is the
eigenvalue of Poincar\'{e}'s problem. Second, this criterion is
generalized to barotropic geophysical flows in $\beta$ plane.
Based on the criteria, the flows are are divided into different
categories of stable flows, which may simplify the further
investigations. And the connections between present criteria and
Arnol'd's nonlinear criteria are discussed. These results extend
the former criteria obtained by Rayleigh, Tollmien and Fj\o rtoft
and would intrigue future research on the mechanism of
hydrodynamic instability.

\end{abstract}
\pacs{47.20.-k, 47.20.Cq, 47.20.Ft, 47.15.Ki } \maketitle

 The stability due to shear in the flow is one of the
fundamental and the most attracting problems in many fields, such
as fluid dynamics, astrophysical fluid dynamics, oceanography,
meteorology et al. The shear instability has been intensively
investigated, which is to the greatly helpful understanding of
other instability mechanisms in complex shear flows. For the
inviscid parallel flow with horizontal velocity profile of $U(y)$,
the general way is to investigate the growth of linear
disturbances by means of normal mode expansion, which leads to the
famous Rayleigh's equation \cite{Rayleigh1880}. Using this
equation, Rayleigh \cite{Rayleigh1880} first proved a necessary
criterion for instability, i.e., Inflection Point Theorem. Then,
Fj{\o}rtoft \cite{Fjortoft1950} found a stronger necessary
criterion for instability. These criteria are well known and have
been applied to understanding the mechanism of hydrodynamic
instability \cite{Drazin1981,Huerre1998,CriminaleBook2003}.
Unfortunately, both criteria are only necessary criteria for
instability, except for some special cases of the symmetrical or
monotone velocity profiles. Tollmien \cite{Tollmien1935} gave a
heuristic result that the criteria are also sufficiency for
instability in these special cases.

The stable criteria also provide a way to categorize the velocity
profiles of the flows. According to Rayleigh's criterion, the
flows are stable if $U''(y)\neq 0$, where $U''(y)$ denotes
$d^2U/dy^2$. And according to Fj{\o}rtoft's criterion, there is
another kind of stable flows if $U''(U-U_s)>0$, where $U_s$ is the
velocity at the inflection point $U''_s=0$. Then if
$U''(U-U_s)<0$, can the flow still be stable? Is there another
kind of stable flows besides the above flows? To answer these
questions, a more restrictive criterion is needed. And the
criterion itself is important for both theoretic researches and
real applications. The aim of this letter is to obtain such a
stability criterion. and other instabilities may be understood via
the investigation here.

For this purpose, Rayleigh's equation for an inviscid parallel
flow is employed
\cite{Rayleigh1880,Drazin1981,Huerre1998,SchmidBook2000,CriminaleBook2003}.
For a parallel flow with mean velocity $U(y)$, the streamfunction
of the disturbance expands as a series of waves (normal modes)
with real wavenumber $k$ and complex frequency
$\omega=\omega_r+i\omega_i$, where $\omega_i$ denotes the grow
rate of the waves. The flow is unstable if and only if
$\omega_i>0$. We study the stability of the disturbances by
investigating the growth rate of the waves, this method is known
as normal mode method. The amplitude of waves, namely $\phi$,
satisfies
 \begin{equation}
 (\phi''-k^2 \phi)-\frac{U''}{(U-c)}\phi=0,
 \label{Eq:stable_parallelflow_RayleighEq}
 \end{equation}
where $c=\omega/k=c_r+ic_i$ is the complex phase speed. The real
part of complex phase speed $c_r=\omega_r/k$ is the wave phase
speed. In fact, Rayleigh's equation is the vorticity equation of
the disturbance \cite{Drazin1981,Huerre1998}. This equation is to
be solved subject to homogeneous boundary conditions
\begin{equation}
\phi=0 \,\, at\,\, y=a,b.
\label{Eq:stable_parallelflow_RayleighBc}
\end{equation}
There are three main categories of boundaries: (i) enclosed
channels with both $a$ and $b$ being finite, (ii) boundary layer
with  either $a$ or $b$ being infinite, and (iii) free shear flows
with  both $a$ and $b$ being infinite.

%Though distinguishing the boundary categories make no difference
%for our following proofs, it does impact the criterion so as to
%the applications.

It is obvious that the criterion for stability is $\omega_i=0$
($c_i=0$), for that the complex conjugate quantities $\phi^*$ and
$c^*$ are also a physical solution of
Eq.(\ref{Eq:stable_parallelflow_RayleighEq}) and
Eq.(\ref{Eq:stable_parallelflow_RayleighBc}). Multiplying
Eq.(\ref{Eq:stable_parallelflow_RayleighEq}) by the complex
conjugate $\phi^{*}$  and integrating over the domain $a\leq y
\leq b$, we get the following equations
\begin{equation}
%\begin{array}{l}
\displaystyle\int_{a}^{b}
[(\|\phi'\|^2+k^2\|\phi\|^2)+\frac{U''(U-c_r)}{\|U-c\|^2}\|\phi\|^2]\,
dy=0%\\
%\displaystyle c_i\int_{a}^{b} \frac{U''}{\|U-c\|^2}\|\phi\|^2\,dy=0
\label{Eq:stable_parallelflow_Rayleigh_Int_Rea}
% \end{array}
 \end{equation}
and
\begin{equation}
%\begin{array}{l}
\displaystyle c_i\int_{a}^{b}
\frac{U''}{\|U-c\|^2}\|\phi\|^2\,dy=0.
\label{Eq:stable_parallelflow_Rayleigh_Int_Img}
% \end{array}
 \end{equation}
Rayleigh  used only
Eq.(\ref{Eq:stable_parallelflow_Rayleigh_Int_Img}) to prove his
theorem. Fj\o rtoft noted that
Eq.(\ref{Eq:stable_parallelflow_Rayleigh_Int_Rea}) should also be
satisfied, then he obtained his necessary criterion. To find a
more sufficient criterion, we shall investigate the conditions for
$c_i=0$. Unlike the former investigations, we consider this
problem in a totally different way: if the velocity profile is
stable ($c_i=0$), then the hypothesis $c_i\neq0$ should result in
contradictions in some cases. Following this, some more
restrictive criteria can be obtained.

To find a stronger criterion, we need to estimate the ratio of
$\int_{a}^{b} \|\phi'\|^2 dy$ to $\int_{a}^{b} \|\phi\|^2 dy$.
This is known as Poincar\'{e}'s problem:
\begin{equation}
\int_{a}^{b}\|\phi'\|^2 dy=\mu\int_{a}^{b}\|\phi\|^2 dy,
\label{Eq:stable_parallelflow_Poincare}
\end{equation}
where the eigenvalue $\mu$ is  positive definition for any $\phi
\neq 0$. The smallest eigenvalue value, namely $\mu_1$, can be
estimated as $\mu_1>(\frac{\pi}{b-a})^2$, like Tollmien
\cite{Tollmien1935} did.

Then using Poincar\'{e}'s relation
Eq.(\ref{Eq:stable_parallelflow_Poincare}), a new stability
criterion may be found: the flow is stable if
$-\mu_1<\frac{U''}{U-U_s}<0$ everywhere.

To get this criterion, we introduce an auxiliary function
$f(y)=\frac{U''}{U-U_s}$, where $f(y)$ is finite at the inflection
point. We will prove the criterion by two steps. At first, we
prove proposition 1: if the velocity profile is subject to
$-\mu_1<f(y)<0$, then $c_r\neq U_s$.

\iffalse Proof: Otherwise, $f(y)$
%
\begin{equation}
   -\mu_1<\frac{U''}{U-U_s}=\frac{U''(U-U_s)}{(U-U_s)^2}\leq\frac{U''(U-U_s)}{(U-U_s)^2+c_i^2},
\end{equation}
%
\fi
%
Proof: Since $-\mu_1<f(y)<0$, then
\begin{equation}
  -\mu_1<\frac{U''}{U-U_s}=\frac{U''(U-U_s)}{(U-U_s)^2}\leq\frac{U''(U-U_s)}{(U-U_s)^2+c_i^2}.
\label{Eq:stable_parallelflow_Rayleigh_inequ}
\end{equation}
Substitution of $c_r=U_s$ and
Eq.(\ref{Eq:stable_parallelflow_Rayleigh_inequ}) into
Eq.(\ref{Eq:stable_parallelflow_Rayleigh_Int_Rea}) results in
\begin{equation}
%\begin{array}{rl}
\displaystyle\int_a^b
[\|\phi'\|^2+k^2\|\phi\|^2+\frac{U''(U-U_s)}{\|U-c\|^2}\|\phi\|^2]\,
dy > 0.
%\displaystyle\int_a^b [(\mu_1+\frac{U''(U-U_s)}{\|U-c\|^2})
%\|\phi\|^2]&>0.
%\end{array}
\end{equation}
This contradicts
Eq.(\ref{Eq:stable_parallelflow_Rayleigh_Int_Rea}). So proposition
1 is proved.

Then, we prove proposition 2: if $-\mu_1<f(y)<0$ and $c_r\neq
U_s$, there must be $c_i^2=0$.

Proof: If $c_i^2\neq0$, then multiplying
Eq.(\ref{Eq:stable_parallelflow_Rayleigh_Int_Img}) by
$(c_r-U_t)/c_i$, where the arbitrary real constant $U_t$ does not
depend on $y$, and adding the result to
Eq.(\ref{Eq:stable_parallelflow_Rayleigh_Int_Rea}), it satisfies
\begin{equation}
%\begin{array}{rl}
\displaystyle\int_a^b
[(\|\phi'\|^2+k^2\|\phi\|^2)+\frac{U''(U-U_t)}{\|U-c\|^2}\|\phi\|^2]\,
dy=0.
%\end{array}
\label{Eq:stable_parallelflow_Sun_Int} \end{equation}
But the above Eq.(\ref{Eq:stable_parallelflow_Sun_Int}) can not
hold for some special $U_t$. For example, let $U_t=2c_r-U_s$, then
there is $(U-U_s)(U-U_t)<\|U-c\|^2$, and
 \begin{equation}
\frac{U''(U-U_t)}{\|U-c\|^2}=
f(y)\frac{(U-U_s)(U-U_t)}{\|U-c\|^2}>-\mu_1.
\label{Eq:stable_parallelflow_Sun_Ust}
 \end{equation}
This yields
\begin{equation} \int_a^b
\{\|\phi'\|^2+[k^2+\frac{U''(U-U_t)}{\|U-c\|^2}]\|\phi\|^2\} dy>0,
\end{equation}
which also contradicts Eq.(\ref{Eq:stable_parallelflow_Sun_Int}).
So proposition 2 is also proved.

Using 'proposition 1: if $-\mu_1<f(y)<0$ then $c_r\neq U_s$' and
'proposition 2: if $-\mu_1<f(y)<0$ and $c_r\neq U_s$ then $c_i =
0$', we find a stability criterion. If the velocity profile
satisfies $-\mu_1<\frac{U''}{U-U_s}<0$  everywhere in the flow, it
is stable. Moreover, the above proof is still valid for $0<f(y)$,
which is equivalent to Fj\o rtoft's criterion. Thus we have the
following theorem.

Theorem 1: If the velocity profile satisfies either
$-\mu_1<\frac{U''}{U-U_s}<0$ or $0<\frac{U''}{U-U_s}$, the flow is
stable.

\begin{figure}
  \includegraphics[width=6cm]{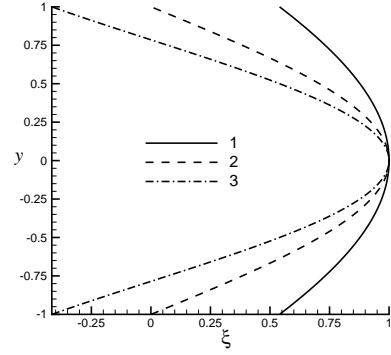}
\caption{Vorticity profiles within the interval $-1\leq y\leq 1$.
Profile 2 ( $\xi=\cos(\pi y/2)$, dashed) is neutrally stable,
while profile 1 ($\xi=\cos( y)$, solid) and profile 3
($\xi=\cos(2y)$, dash doted) are stable and unstable,
respectively. } \label{Fig:vorticity_profile}
\end{figure}

This criterion is more restrictive than Fj\o rtoft's criterion. As
known from Fj\o rtoft's criterion, the necessary condition for
instability is that the base vorticity $\xi=-U'$ has a local
maximal in the profile. Noting that $U''/(U-U_s)\approx
\xi_s''/\xi_s$ near the inflection point, where $\xi_s$ is the
vorticity at inflection point, it means that the base vorticity
$\xi$ must be convex enough near the local maximum for
instability, i.e., the vorticity should be concentrated somewhere
in the flow for instability. A simple example can be given by
following Tollmien's way \cite{Tollmien1935}. As shown in
Fig.\ref{Fig:vorticity_profile}, there are three vorticity
profiles within the interval $-1\leq y\leq 1$, which have local
maximal at $y=0$. Profile 2 ($U=-2\sin(\pi y/2)/\pi$) is neutrally
stable, while profile 1 ($U=-\sin(y)$) and profile 3
($U=-\sin(2y)/2$) are stable and unstable, respectively.

Moreover, the stabile criterion for the parallel inviscid flows
can be applied to the barotropic geophysical flows in $\beta$
plane, like Kuo did \cite{KuoHL1949}. This is a generalized stable
criterion, we state it as a new theorem.

Theorem 2: The flow is stable, if the velocity profile  satisfies
either $-\mu_1<\frac{U''-\beta}{U-U_s}<0$ or
$0<\frac{U''-\beta}{U-U_s}$ in the flow, where $U_s$ is the
velocity at the point $U''=\beta$.

The criteria proved above may shed light on the investigation of
vortex dynamics. Both Theorem 1 and
Fig.\ref{Fig:vorticity_profile} show that it is the vorticity
profile rather than the velocity profile that dominates the
stability of the flow. This means that the distribution of
vorticity dominates the shear instability in parallel inviscid
flow, which is essential to understanding the role of vorticity in
fluid. So an unstable flow might be controlled just by adjusting
the vorticity distribution according to present results. This is
an very fascinating problem, but can not be discussed in detail
here.

\begin{figure}
  \includegraphics[width=6cm]{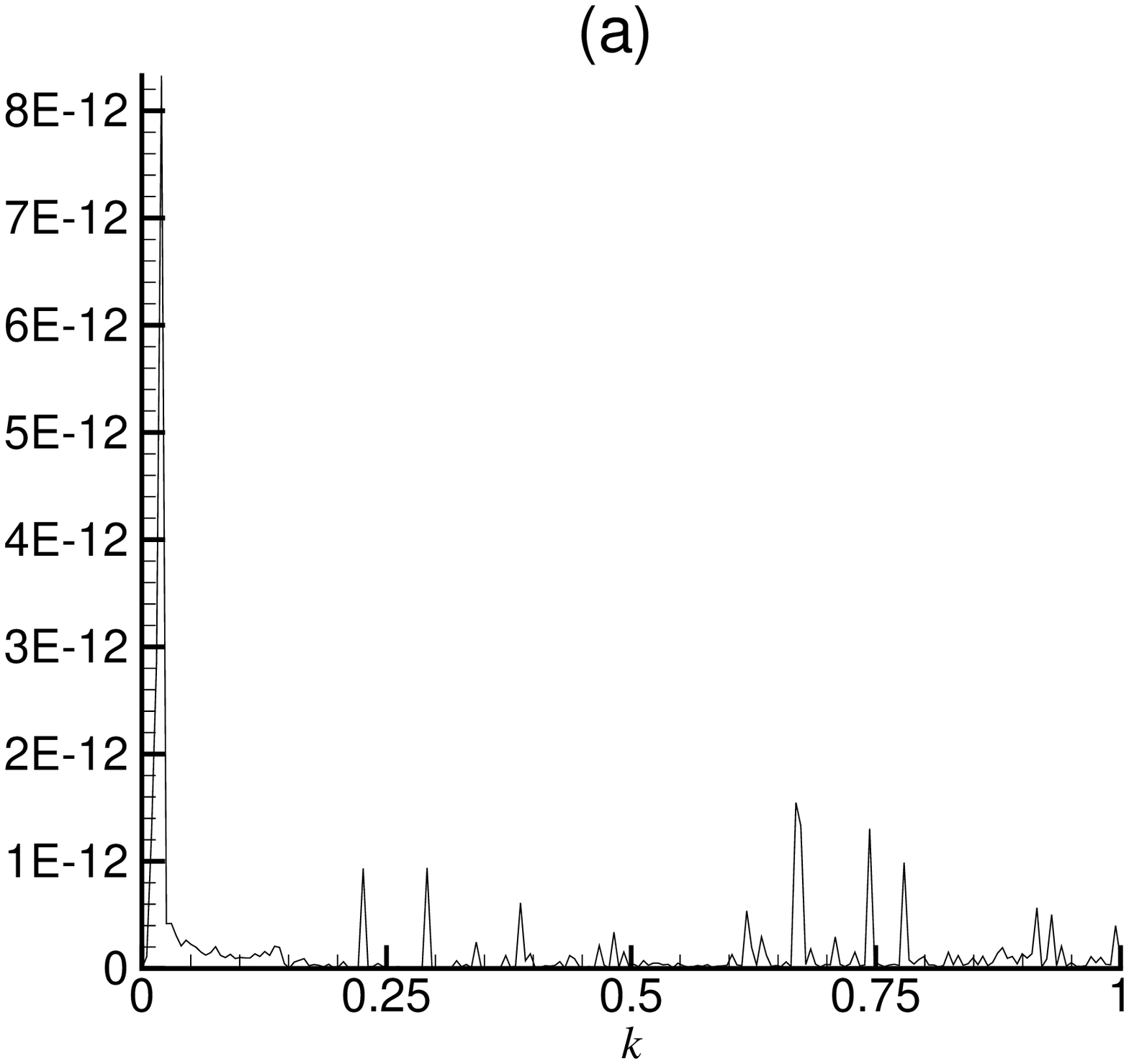}
  \includegraphics[width=6cm]{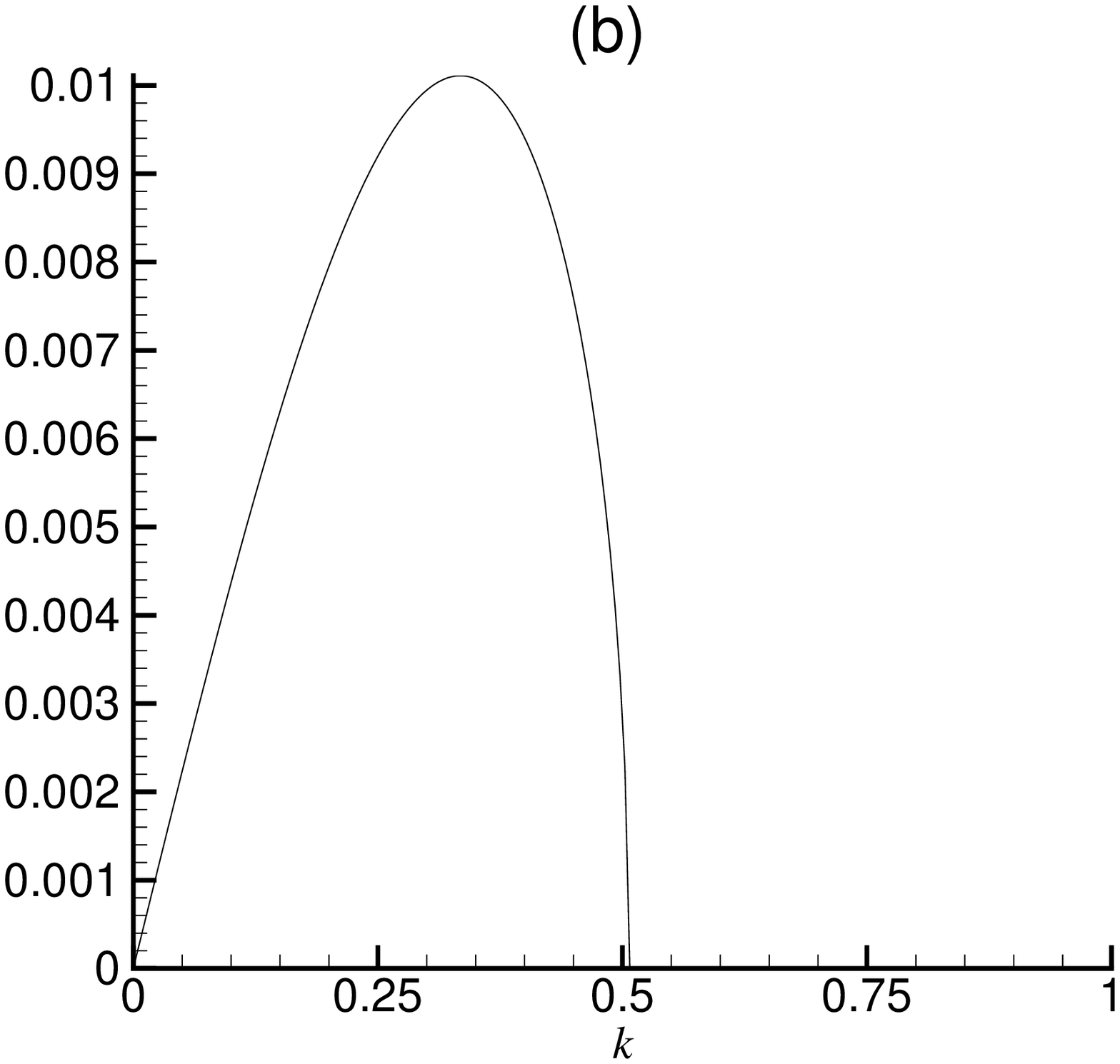}
\caption{Growth rate $\omega_i$ as an function of wavenumber $k$,
(a) for $U_1=\tanh(1.1 y)$, (b) for $U_2=\tanh(1.3 y)$, both
within the interval $-1\leq y\leq 1$. } \label{Fig:Growth}
\end{figure}

To show the power of the criteria obtained above, we consider the
stability of velocity profile $U=\tanh(\alpha y)$ within the
interval $-1\leq y\leq 1$, where $\alpha$ is a constant. This
velocity profile is an classical model of mixing layer, and has
been investigated by many researchers (see
\cite{Huerre1998,SchmidBook2000,CriminaleBook2003} and references
therein). Since $U''(U-U_s)=-2\alpha^2\tanh^2(\alpha
y)/\cosh^2(\alpha y) <0$ for $-1\leq y\leq 1$, it might be
unstable for any $\alpha$ according to both Rayleigh's and Fj\o
rtoft's criteria. But it can be derived from Theorem 1 that the
flow is stable for $\alpha^2<\pi^2/8 $. For example, we choose
$\alpha_1=1.1$ and $\alpha_2=1.3$ for velocity profiles $U_1(y)$
and $U_2(y)$. The growth rate of the profiles can be obtained by
Chebyshev spectral collocation method \cite{SchmidBook2000} with
100 collocation points, as shown in Fig.\ref{Fig:Growth}. It is
obvious that $c_i=0$ for $U_1$ and $c_i>0$ for $U_2$, which agrees
well with the criteria obtained above. This is also a
counterexample that Fj\o rtoft's criterion is not sufficient for
instability. So this new criterion for stability is more useful in
real applications.
%
%This implies the future works to understanding the interaction
%between waves and background flow from energy transport point of
%view.

The present stable criteria give a affirmative answer to the
questions at the beginning, i.e., there are some stable flows if
$U''(U-U_s)<0$. Based on the former criteria, the velocity
profiles can be categorized as follows: (\romannumeral1) without
inflection point (Reyleigh's criterion), (\romannumeral2) $f(y)>0$
(Fj{\o}rtoft's criterion), and (\romannumeral3) $\mu_1<f(y)<0$
(present criterion). Then the flow might be unstable only for
$f(y)<\mu_1$ and $f(y)$ changing sign within the interval.
However, if $f(y)$ changes sign somewhere within the interval
$[a,b]$, then the flow is stable. For that $f(y)$ changing sign
implies $U'''_s=0$ but $U''''_s\neq0$, so $U''$ does not change
sign near the inflection point. Thus $c_i$ must vanish in
Eq.(\ref{Eq:stable_parallelflow_Rayleigh_Int_Img}), i.e., the flow
is stable for $f(y)$ changing sign within the interval. In this
way, the flow might be unstable only for $f(y)<\mu_1$ somewhere,
which will intrigue further studies on this problem. In fact,
there are still stable flows if $\mu_1<f(y)$ is violated.

Recall the proof of theorem 1, it is found that the following
Rayleigh's quotient $I(f)$ plays a key role in determination the
stability of the flows.
\begin{equation}
I(f)=\min_{\phi} \frac{\int_{a}^{b}
[\,\|\phi'\|^2+f(y)\|\phi\|^2\,]\, dy}{\int_{a}^{b} \|\phi\|^2}
\label{Eq:stable_paralleflow_sun_Energy}
\end{equation}
Noting that the proof of theorem 1 is still valid in the case of
$I(f)>0$. We have such result: the flows are stable if $I(f)>0$.
Though this criterion is more restrictive than that in theorem 1,
it is inconvenient for the real applications due to unknown value
of Rayleigh's quotient $I(f)$. Theorem 1 is more convenient for
the real applications in different research fields.

The idea of categorization the velocity profiles of the flows may
simplify the investigation of stability problem. It can be seen
from Rayleigh's equation
Eq.(\ref{Eq:stable_parallelflow_RayleighEq}) that the stability of
profile $U(y)$ is not only Galilean invariant, but also
independent from the the magnitude of $U(y)$ due to linearity. So
the stability of $U(y)$ is the same as that of $AU(y)+B$, where
$A$ and $B$ are any arbitrary nonzero real numbers. As the value
of $U''(U-U_s)$ in Fj\o rtoft's criterion is only Galilean
invariant but not magnitude free, it satisfies only part of the
Rayleigh's equation's properties. On the other hand the value of
$U''/(U-U_s)$ satisfies both conditions, this is the reason why
the criteria in both Arnol'd's theorems and present theorems are
the functions of $U''/(U-U_s)$. Since the stability of inviscid
parallel flow depends only on the velocity profile's geometry
shape, namely $f(y)$, and the magnitude of the velocity profile
can be free, then the instability of inviscid parallel flow could
be called "geometry shape instability" of the velocity profile. As
the above investigation shows that the inviscid shear instability
is only associated with the geometry of velocity profile. The
concept of "geometry shape instability" would be help in further
investigations. This distinguishes from the viscous instability,
which is also associated with the magnitude of the velocity
profile.

As mentioned above, we have investigated the stability of the
flows via Rayleigh's equation, while Arnol'd \cite{Arnold1969}
considered the hydrodynamic stability in a totally different way.
He studied the conservation law of the inviscid flow via Euler's
equations and found two nonlinear stability conditions by means of
variational principle. So what is the relationship between the
linear criteria and the nonlinear ones?

It is very interesting that the linear stability criteria match
Arnol'd's nonlinear stability theorems very well. Applying
Arnol'd's First Stability Theorem to parallel flow, the stable
criterion is $0<C_1<(U-U_s)/U''<C_2<\infty$ everywhere in the
flow, where $C_1$ and $C_2$ are constants. This corresponds to
Fj\o rtoft's criterion for linear stability, and is well known
\cite{Drazin1981,Dowling1995}. Here we find that Theorem 1 proved
above corresponds to Arnol'd's Second Stability Theorem, i.e., the
stable criterion is $0<C_1<-(U-U_s)/U''<C_2<\infty$ everywhere in
the flow. Given $C_1=1/\mu_1$, Arnol'd's Second Stability Theorem
is equivalent to Theorem 1. Moreover, the proofs here are similar
to Arnol'd's variational principle method. For the arbitrary real
number $U_t$, which is like a Lagrange multiplier in variational
principle method, plays a key role in the proofs. So that the
above Theorem 1 is similar to Arnol'd's theorems.

Unfortunately, Arnol'd's nonlinear stability theorems, though
quite useful in the geophysical flows \cite{Dowling1995}, are
seldom known by the scientists in other fields. The main reason is
that the proofs of Arnol'd's theorems are very advanced and
complex in mathematics for most general scientists in different
fields to understand. Although Dowling \cite{Dowling1995}
suggested that Arnol'd's idea need to be added to the general
fluid-dynamics curriculum, his suggestion has not been followed
even 10 years later. Compare with  Arnol'd's theorems, the
theorems proved here are equivalent in some sense but much simpler
and easier to understand, therefore it is more convenient to use
our new results in applications.

In summary, the general stability criteria are obtained for
inviscid parallel flow. These results, which are equivalent to
Arnol'd's nonlinear theorems, extend the former theorems proved by
Rayleigh, Tollmien and Fj\o rtoft. Based on the criteria, the
velocity profiles are divided into different categories, which may
simplify the further investigations. In general, these criteria
would intrigue future research on the mechanism of hydrodynamic
instability and to understand the mechanism of turbulence. And it
also sheds light on the flow control and investigation of the
vortex dynamics.

The author thanks Prof. Sun D-J at USTC, Dr. Yue P-T at UBC
(Canada) and two anonymous referees for their useful comments.
This work was original from author's dream of understanding the
mechanism of instability in the year 2000, when the author was a
graduated student and learned the course of hydrodynamic stability
by Prof. Yin X-Y at USTC.

\iffalse

\bibliography{MSH1}

\fi

\end{document}